\documentclass[acmsmall,screen]{acmart}
\renewcommand\footnotetextcopyrightpermission[1]{}
\AtBeginDocument{%
  \providecommand\BibTeX{{%
    \normalfont B\kern-0.5em{\scshape i\kern-0.25em b}\kern-0.8em\TeX}}}

\setcopyright{rightsretained}
\acmPrice{}
\acmDOI{10.1145/3428201}
\acmYear{2020}
\copyrightyear{2020}
\acmSubmissionID{oopsla20main-p24-p}
\acmJournal{PACMPL}
\acmVolume{4}
\acmNumber{OOPSLA}
\acmArticle{133}
\acmMonth{11}

\usepackage{booktabs}
\usepackage{subcaption}
\usepackage{braket}
\usepackage{listings}
\usepackage{enumerate}
\usepackage{xcolor}
\usepackage{graphics}
\usepackage{subcaption}
\usepackage{mathtools}
\usepackage{dsfont}
\usepackage{epstopdf}
\usepackage{hyphenat}
\usepackage{tikz}
\usepackage{mathrsfs}
\usepackage{enumitem}
\usepackage{microtype}

\usetikzlibrary{shadows.blur}
\usetikzlibrary{shapes.symbols}

\definecolor{hl}{rgb}{0.2,0.2,0.9}
\newcommand{\HL}[1]{\textcolor{hl}{#1}}

\begin{document}
\newcommand{\todo}[1]{\textcolor{red}{TODO: #1}}
\graphicspath{{figures/}}

\title{Assertion-Based Optimization of Quantum Programs}

\author{Thomas H\"aner}
\orcid{0000-0002-4297-7878}
\affiliation{
  \institution{ETH Z\"urich}
  \country{Switzerland}
}
\additionalaffiliation{Microsoft Quantum, Z\"urich}

\author{Torsten Hoefler}
\orcid{0000-0001-9611-7171}
\affiliation{
  \institution{ETH Z\"urich}
  \country{Switzerland}
}

\author{Matthias Troyer}
\orcid{0000-0002-1469-9444}
\affiliation{
  \institution{Microsoft}
  \country{USA}
}

\begin{abstract}
Quantum computers promise to perform certain computations exponentially faster than any classical device. Precise control over their physical implementation and proper shielding from unwanted interactions with the environment become more difficult as the space/time volume of the computation grows. Code optimization is thus crucial in order to reduce resource requirements to the greatest extent possible. Besides manual optimization, previous work has adapted classical methods such as constant-folding and common subexpression elimination to the quantum domain. However, such classically-inspired methods fail to exploit certain optimization opportunities across subroutine boundaries, limiting the effectiveness of software reuse. To address this insufficiency, we introduce an optimization methodology which employs annotations that describe how subsystems are entangled in order to exploit these optimization opportunities. We formalize our approach, prove its correctness, and present benchmarks: Without any prior manual optimization, our methodology is able to reduce, e.g., the qubit requirements of a 64-bit floating-point subroutine by $34\times$.
\end{abstract}



\maketitle

\section{Introduction}

Quantum computers promise to solve certain computational tasks exponentially faster than classical computers. As a result, significant resources are being spent in order to make quantum computing become reality.
In anticipation of the first quantum computers, the most promising applications are being identified and manually optimized for specific problems of practical interest, resulting in great resource savings~\cite{reiher2016,hastings2015improving,kutin2006shor,roetteler2017factoring,gidney2019factor}.
Such improvements are crucial, given the considerable overhead due to quantum error correction~\cite{fowler2012surface} and the difficulty of engineering large-scale quantum computers.

To identify promising applications for quantum computing, it is necessary to develop a detailed understanding of all components of the quantum algorithm being studied. One possible approach is to implement the algorithm in a quantum programming language, as this also enables testing and debugging.
To this end, a host of software packages, programming languages, methodologies, and compilers for quantum computing have been developed~\cite{chong2017programming,steiger2018projectq,pyquil,javadiabhari2014scaffcc,green13,Svore:2018:QES:3183895.3183901,qiskit,paykin2017qwire}. In addition to providing the necessary layers of abstraction to facilitate software development, these packages include optimizing compilers that aim to reduce width and depth of the resulting quantum circuit, e.g., by merging operations at various layers of abstraction~\cite{haner2018software}. Further optimization opportunities can be created by employing a set of commutation relations~\cite{nam2018automated} to reorder operations. Moreover, several methods have been developed for exact circuit synthesis with certain optimality guarantees~\cite{grosse2007exact,grosse2009exact,amy2013meet,KMM1,meuli2018sat}. While these methods alone are not suitable for optimization of large-scale quantum circuits, they can be combined with heuristics~\cite{amy14polynomial,nam2018automated}.

Despite these efforts, most of the progress made in, e.g., quantum chemistry has been due to manual optimization~\cite{jones2012faster,hastings2015improving}. This suggests that the capabilities of optimizing compilers may still be significantly improved; especially at higher levels of abstraction where manual optimizations are carried out today.
Such optimization opportunities usually arise when two or more existing subroutines are combined instead of directly implementing the desired functionality more efficiently. Exploiting these opportunities is only possible when optimizing across the boundaries of both quantum and classical subroutines. To enable such transformations, an optimizing compiler requires access to additional information such as the circumstances under which a given subroutine is invoked.
While compilers may not be able to infer the semantics of a given program and its subroutines, such information may be extracted from assertions, and then used for program \textit{optimization}. Specifically, we propose to use the pre- and postconditions of each subroutine in order to gather information about the state of the quantum computer before and after completion of the subroutine. Fig.~\ref{fig:hlinfo} depicts a mixed quantum/classical code example and annotations of postconditions and derived facts for both classical and quantum subroutines.
The gathered information may then be combined, e.g., with conditions specifying the circumstances under which a given subroutine acts trivially. If these conditions are met, our methodology removes such operations and is thus able to reduce both space and time requirements of a given quantum program.

\definecolor{lgrey}{rgb}{0.96,0.96,0.96}
\definecolor{postccolor}{rgb}{1,0.88,0.88}
\definecolor{postccolordark}{rgb}{1,0.75,0.75}
\begin{figure}[t]
\centering
\begin{tikzpicture}[]
\tikzstyle{postcstyle}=[blur shadow];
\draw[draw=none,rounded corners,fill=lgrey] (-2.8,0.6) rectangle (3,-3.5);

\begin{scope}[yshift=0cm]
\draw[draw=none,rounded corners=1pt, ultra thin,fill=hl!50!white] (-2.8,1) rectangle (-2.6,0.8);
\draw[draw=none,rounded corners=1pt, ultra thin,fill=black] (-2.8,1.4) rectangle (-2.6,1.2);
\node[text width=3cm] at (-0.9,1.3) {\small Quantum logic};
\node[text width=3cm] at (-0.9,0.88) {\small \HL{Classical logic}};
\end{scope}

\node[text width=3cm] at (4.9,0.9) {\small Extracted information:};

\node[align=left, text width=5cm] at (0,0.25) {\HL{\texttt{{\bfseries qureg} reg[n];}}};
\node[align=left, text width=5cm] at (0.3,-0.4) {\textit{[updates to {reg}]}};
\node[align=left, text width=5cm] at (0.1,-0.35) {\rotatebox[]{90}{\textit{$\cdots$}}};
\node[align=left, text width=5cm] at (0,-1) {\texttt{{\bfseries \HL{while} Measure}(q) \HL{== 0:}}};

\node[align=left, text width=5cm] at (0.5,-1.75) {\texttt{\HL{{\bfseries set} pos =} find\_first1(reg);}};

\node[align=left, text width=5cm] at (0.5,-2.5) {\texttt{shift(reg, pos);}};
\node[align=left, text width=5cm] at (0.8,-3.05) {\rotatebox[]{90}{$\cdots$}};

\node[rectangle, draw=postccolor, fill=postccolor, rounded corners, align=left, text width=5cm, text height=0.4cm, label=center:\text{$\forall i\in\{0,...,n-1\}:\texttt{reg}[i] = \ket0$}, postcstyle] (declpost) at (6,0.25) {};
\node[draw=postccolordark,circle,fill=postccolordark,inner sep=1.2pt] (decl) at (0,0.25) {};
\draw[postccolordark] (decl)--(declpost);

\node[rectangle, draw=postccolor, fill=postccolor, rounded corners, align=left, text width=1.5cm, text height=0.4cm, label=center:\text{$\texttt{q} = \ket0$}, postcstyle] (whilepost) at (4.25,-1) {};
\node[draw=postccolordark,circle,fill=postccolordark,inner sep=1.2pt] (while) at (1.4,-1) {};
\draw[postccolordark] (while)--(whilepost);

\node[rectangle, draw=postccolor, fill=postccolor, rounded corners, align=left, text width=3.5cm, text height=0.4cm, label=center:\text{$\forall i < \texttt{pos}: \texttt{reg}[i] = \ket 0$}, postcstyle] (first1post) at (5.25,-1.75) {};
\node[draw=postccolordark,circle,fill=postccolordark,inner sep=1.2pt] (first1) at (2.8,-1.75) {};
\draw[postccolordark] (first1)--(first1post);

\node[rectangle, draw=postccolor, fill=postccolor, rounded corners, align=left, text width=2cm, text height=0.4cm, label=center:\text{$\texttt{reg}[0]\neq \ket 0$}, postcstyle] (first1post) at (4.5,-2.5) {};
\node[draw=postccolordark,circle,fill=postccolordark,inner sep=1.2pt] (first1) at (1,-2.5) {};
\draw[postccolordark] (first1)--(first1post);

\end{tikzpicture}
\caption{Example depicting a quantum program and the corresponding information that our methodology infers from the postconditions of both classical and quantum subroutines. It uses this information to exploit optimization opportunities that arise due to subroutine composition. Here, the \texttt{shift} subroutine may be optimized, as discussed in Section~\ref{sec:examples}.}
\label{fig:hlinfo}
\end{figure}

\paragraph{Contributions.} We develop a formalism that allows us to express (1) how subsystems (subsets of qubits) of the quantum computer are entangled and (2) under what circumstances the program may be optimized. Our methodology then uses this information for quantum program optimization. We prove its correctness and present a prototype implementation that successfully reduces the quantum resource requirements of common arithmetic subroutines by up to $410\times$ (for a 64-bit floating-point subroutine). The chosen subroutines are frequently used in a wide range of quantum algorithms, including Shor's algorithm for factoring~\cite{shor94}, HHL for solving linear systems of equations~\cite{harrow2009quantum}, algorithms for quantum chemistry~\cite{babbush2016exponentially}, and Grover's algorithm~\cite{grover1996fast} when used for optimization. Our proposed automatic program-level optimizations are thus beneficial for a wide range of quantum computing applications.

\paragraph{Related work.} By carrying out optimizations across different subroutines of the quantum program, our methodology complements available tools for lower-level circuit optimization and synthesis~\cite{amy14polynomial,amy2013meet,nam2018automated}. Crucially, our methodology enables more effective use of abstractions when implementing libraries for quantum computing because it is able to remove the resulting overheads. Furthermore, previous work on high-level quantum program optimization~\cite{steiger2018projectq,haner2018software}, which adapted common subexpression elimination and constant-folding to the quantum domain, cannot handle the examples we present in Section~\ref{sec:examples}. Moreover, the verification of quantum programs has been addressed through the introduction of a quantum Hoare logic~\cite{Ying:2012:FLQ:2049706.2049708}. In contrast, our methodology uses the information that it gathers from pre- and postconditions of subroutines for optimization purposes.

\section{Preliminaries}

In this section, we provide some background on quantum computing and quantum programs. For a more in-depth treatment of quantum computing, we refer to the textbook by Nielsen and Chuang~\cite{nielsen2010}.

\subsection{Qubits and Gates}

Whereas classical computers manipulate bits in order to solve a certain computational task, their quantum counterparts operate on so-called quantum bits, or \textit{qubits}. A qubit is a two-level quantum system, i.e., a system which can be in two distinguishable states. An example would be the ground and first excited state of an ion, where we can denote the ground state by 0 and the first excited state as 1.

The principle of \textit{quantum superposition} states that a single qubit can be in a complex superposition of its two levels. This means that there are two complex numbers associated with the quantum state of a qubit: the contribution from the 0-state and another one from the 1-state. Let us denote these two complex numbers by $\alpha_0$ and $\alpha_1$, respectively, where we require that $|\alpha_0|^2+|\alpha_1|^2=1$. Given a qubit in a quantum state which is described by these two values, the probability of observing the qubit in state 0 or 1 is $|\alpha_0|^2$ or $|\alpha_1|^2$, respectively. Note that the normalization condition above ensures that the two probabilities sum up to 1.

If we add a single qubit to a system of $n-1$ qubits, the resulting system must be described in general using twice as many complex values due to \textit{quantum entanglement}. Two subsystems being entangled means that one cannot write down the state of the entire system as a product state of the two subsystems. As a result, operations that act on one subsystem (such as measurement) may have nontrivial effects on the other. The state of an $n$-qubit quantum computer can be described using $2^n$ complex amplitudes that correspond to the contributions stemming from the all-zero state, the all-zero state but where the last bit is 1, up to the all-one state. We denote the corresponding amplitudes by $\alpha_{0\cdots 00},\alpha_{0\cdots 01},...,\alpha_{1\cdots11}$ and, as a simplification, we interpret the indices as a number written in binary format, allowing to write $\alpha_0, \alpha_1,...,\alpha_{2^n-1}$. When reading out, or \textit{measuring}, all $n$ qubits, the probability of observing the binary representation of $i$ is given by $|\alpha_i|^2$. Once observed, the entire quantum system collapses onto the observed outcome, meaning that $\alpha_i=1$ and $\alpha_j=0$ for all $j\neq i$. In particular, repeated measurement results in the same answer.

When dealing with quantum systems, one typically employs the so-called \textit{Dirac notation}, where state vectors correspond to so-called \textit{kets} that are denoted by $\ket{\cdot}$, e.g., $\ket{\psi}$.
Continuing the example from above, let $\ket{\psi}$ denote the quantum state of an $n$-qubit quantum computer. Using $\ket i$ with $i\in\{0,1,...,2^n-1\}$, we can write
\begin{equation*}
	\ket{\psi}=\sum_{i=0}^{2^n-1}\alpha_i\ket{i}\;,
\end{equation*}
with $\alpha_i\in\mathds C$ the contribution from the $\ket i$-state and $\sum_i|\alpha_i|^2=1$. As above, the binary representation of $i$ lets us determine the value of each of the $n$ qubits in the $i$-th basis state $\ket i$. We note that the set $\{\ket i, i\in\{0,...,2^n-1\}\}$ is called the \textit{computational basis} in the quantum computing literature.

Due to the relationship between amplitudes and probabilities, all operations on qubits, so-called \textit{quantum gates}, must preserve inner-products. As a result, quantum gates must be \textit{unitary}, which means that for a quantum gate $U$,
\begin{equation*}
	U^\dagger U = UU^\dagger=\mathds 1\;,
\end{equation*}
where $U^\dagger$ denotes the Hermitian adjoint of $U$. Note that this also implies that all quantum gates must be reversible and, therefore, that only reversible operations can be implemented using quantum gates.
Such unitary operations may also be applied \textit{controlled} on another qubit, meaning that they are applied if the control qubit is 1. Formally, the controlled version of $U$ is
\begin{equation*}
	U^c := \ket0\bra0\otimes\mathds 1 + \ket1\bra1\otimes U\;,
\end{equation*}
where $\ket c\bra c$ is the projector onto the subspace in which the control qubit has the value $c\in\{0,1\}$ and $\otimes$ denotes the tensor product. Since the control qubit may be in a superposition, the state after applying $U^c$ is in a superposition of having and not having applied $U$.

\subsection{Quantum Programs}

\begin{figure}[t]
	\centering
	\includegraphics[width=0.4\linewidth]{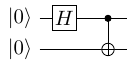}
	\caption{Quantum circuit example: Each line represents a qubit and operations are drawn as boxes (e.g., the Hadamard operation $H$) or other symbols such as the controlled NOT or CNOT, which is depicted as $\oplus$ connected and attached to the filled circle on the control qubit. Time advances from left to right.}
	\label{fig:bellpair}
\end{figure}
A quantum program is a classical program that, in addition to classical instructions, executes so-called \textit{quantum circuits} on a quantum co-processor.
In a quantum circuit, each qubit is represented as a horizontal line and quantum gates are denoted by boxes or other symbols on these lines, with time moving from left to right. See Fig.~\ref{fig:bellpair} for an example. It consists of a Hadamard gate $H$ and a controlled NOT or CNOT gate. The CNOT is drawn as a NOT gate (denoted by $\oplus$) that is connected to a filled circle on the control qubit.

\begin{definition}{Quantum instruction.}{}
Let $O\ket{q_1,...,q_k}$ denote a \textit{quantum instruction}. It consists of an operation $O$ and a $k$-tuple of qubits $(q_1,...,q_k)$, where the operation may be a quantum gate or a classical instruction (allocation, deallocation, measurement).
\end{definition}

Every circuit consists of the following 4 steps:
\begin{enumerate}
\item Allocate $n$ qubits in state $\ket{0}^{\otimes n}:=\ket{0\cdots 0}$ ($n$ zeros)
\item Apply quantum gates to these qubits
\item Measure some or all of the qubits
\item Deallocate measured qubits
\end{enumerate}
Upon completion, the quantum co-processor returns a set of classical bits, the so-called \textit{measurement results}. Depending on these results, the classical processor may then provide further quantum circuits to evaluate in order to solve the computational problem at hand. At the end of the entire quantum program, all qubits are deallocated again.

Since qubits are an extremely scarce resource, it is crucial to keep the number of allocated qubit minimal at any given point throughout the circuit and to deallocate all qubits which are no longer in use. Using the principle of deferred measurement~\cite{nielsen2010}, this means that as soon as the last operation on a given qubit has finished, the qubit can be measured and then freed for further use in the ongoing computation.

\section{Quantum program optimization using assertions}

In this section, we introduce the basic idea of our assertion-based optimization methodology, followed by a proof of its correctness.

\subsection{Entanglement Description Assertions for Quantum Program Optimization}

Our goal is to use the knowledge of how subsystems are entangled for the purpose of quantum program optimization. To this end, we introduce the concept of \textit{entanglement description assertions} that capture this information.

In particular, we introduce a formalism to describe the entanglement between qubits of the quantum computer throughout the execution of the quantum circuit. This entails statements that assert \textit{entanglement descriptions} (to be defined next), that is, statements of the form
\begin{equation*}
	\text{``}\texttt{q} == f(\texttt{q, r})\text{''},\; \text{``}\texttt{q} \geq f(\texttt{q, r})\text{''}, \text{ etc.},
\end{equation*}
where \texttt{q} and \texttt{r} refer to quantum registers and $f$ is a function of two registers returning one register of bits. Since \texttt{q} and \texttt{r} refer to quantum registers, they may be in superposition and entangled with other qubits in the system. The following definition assigns a precise meaning to these \textit{entanglement description assertions} with respect to the state vector of the entire $n$-qubit quantum computer,
\begin{equation*}
	\ket{\psi}=\sum_{i=0}^{2^n-1}\alpha_i\ket i\;.
\end{equation*}

\begin{definition}{Entanglement description assertion.}{}
Let $\square_\text{cmp}$ denote a comparison operator, $f:\{0,1\}^{k}\times \{0,1\}^{m}\rightarrow \{0,1\}^{k}$ a function on $k+m$ bits returning $k$ bits, and let \texttt{q,r} be quantum registers consisting of $k$ and $m$ qubits, respectively.
The \textit{entanglement description assertion} $A(q,r)=\texttt{q }\square_\text{cmp}\,f(\texttt{q, r})$ on the $n$-qubit quantum state $\ket{\psi}$ asserts that
\begin{equation*}
	\forall i\in\{0,...,2^n-1\}:\left(\,|\alpha_i|>0\implies A\left(\mathscr Q(i),\mathscr R(i)\right)\,\right)\;,
\end{equation*}
where the functions $\mathscr Q : \{0,1\}^n\rightarrow \{0,1\}^k$ and $\mathscr R : \{0,1\}^n\rightarrow \{0,1\}^m$ extract the bits corresponding to the quantum registers \texttt{q} and \texttt{r}, respectively, from the $n$-bit index $i$ of the computational basis state $\ket i=\ket{i_{n-1},...,i_0}$.
\end{definition}

With this definition in place, let us revisit the $\ket0$-control qubit example from the previous section and cast it as an \textit{entanglement description assertion}.

\begin{example}{}
To express that a control qubit (denoted by $\ket c$) is in a definite state $\ket0$, let $f(\cdot,\cdot)=0$ and $\square_\text{cmp}$ be the equals comparison operator in the above definition. Then
$A(c, \cdot)=(\texttt{c} == 0)$. For the corresponding state $\ket{\psi}$, this means that $\alpha_i = 0$ whenever $i$ corresponds to a state where the control qubit is 1. As a result, the action of a controlled gate $U^c$ on $\ket{\psi}$ is always trivial.
\end{example}

This shows that such assertions can be used to express knowledge about qubits that are in a definite state. This piece of information can, when combined with classical constant-folding, be used for optimization. However, in order to do so in a more general setting, the optimizer also needs information which specifies the conditions for an operation to be trivial. We call this information \textit{triviality conditions}.

\begin{definition}{Triviality condition.}{}
Let $A(q,r)$ be an entanglement description assertion on the quantum state $\ket{\psi}$. $A(q,r)$ is a \textit{triviality condition} of a quantum operation $U$ if
\begin{equation*}
	A(q,r)\implies U\ket{\psi}=\ket{\psi}\;,
\end{equation*}
meaning that $U$ acts as the identity if $A(q,r)$ is satisfied by $\ket{\psi}$.
\end{definition}

\begin{example}{}
Continuing the $\ket0$-control qubit example, the triviality condition of the controlled unitary $U^c$ would read $\{c==0\}$ and, if this is satisfied as in the previous example, $U^c$ may be removed from the circuit.
\end{example}

Therefore, using these two definitions, we can describe and carry out classical constant-folding. In order to see that this approach is strictly more powerful than classical constant-folding, consider the following example.

\begin{example}{}\label{ex:bell}
Let $\ket{\psi}$ denote the quantum state of a two-qubit quantum computer. Initially, $\ket{\psi}=\ket{00}$ and our quantum program consists of two operations: 1) Prepare a Bell-pair and 2) swap the two qubits by applying a Swap gate.
The Bell-pair preparation routine has $\{q_0 == 0,q_1 == 0\}$ as preconditions and ensures that $\{q_0 == q_1\}$ as a postcondition. In particular, given that the preconditions are satisfied, the Bell-pair preparation circuit in Fig.~\ref{fig:bellpair} transforms the state $\ket{00}$ to
\begin{equation*}
	\frac 1{\sqrt2}(\ket{00}+\ket{11})\;,
\end{equation*}
The amplitudes of this quantum state are $\alpha_{00}= 1/\sqrt 2$, $\alpha_{01}=\alpha_{10}=0$, and $\alpha_{11}=1 / \sqrt 2$. It is easy to check that the postcondition holds, i.e., that
\begin{equation*}
	\forall i\in\{0,1,2,3\}: |\alpha_i|>0\implies \{i_0 == i_1\},
\end{equation*}
where $i_0$ and $i_1$ denote the 0th and 1st bit of $i$, respectively.
Since swaps are trivial if $q_0==q_1$, which is satisfied by the state above, the Swap gate can be removed from the circuit. We note that the same reasoning applies if the Hadamard gate in the Bell-pair preparation circuit is replaced by an arbitrary rotation gate. In this case, $\alpha_{01}=\alpha_{10}=0$ still holds, and so the Swap gate may be removed.
\end{example}

Since the quantum state in the example above is in a superposition, it is clear that regular constant-folding cannot successfully perform this optimization. Using our entanglement description assertions, however, this becomes feasible. This shows that optimization using entanglement description assertions is strictly more powerful than classical constant-folding.

\subsection{Correctness of our Methodology}
We now prove that optimizations based on entanglement description assertions and triviality conditions leave the action of the overall quantum program invariant.

\begin{theorem}[Correctness]
Let $\mathcal C$ be a quantum circuit that gets sent to a perfect\footnote{Since circuit optimization may improve, e.g., success probability for a real device, we assume a hypothetical, perfect device.} quantum device during execution of the quantum program $\mathcal P$. Applying assertion-based optimization to $\mathcal C$ will not change the output of $\mathcal P$.
\end{theorem}
\begin{proof}
Let $A(q,r)$ denote an entanglement description assertion on the state $\ket{\psi}$ of the quantum device at a given point during the execution of $\mathcal C$ and let $U$ be the next subroutine to be executed. Moreover, let $A'(q,r)$ be a triviality condition of $U$ such that
\begin{equation}\label{eq:AimplA'}
A(q,r)\implies A'(q,r).
\end{equation}
Then, applying $U$ is equivalent to
\begin{align*}
\ket{\psi}\mapsto U\ket{\psi} &= \sum_{i=0}^{2^n-1} \alpha_i U\ket i
\\&= \sum_{i : A(\mathscr Q(i), \mathscr R(i))} \alpha_i U\ket i + \sum_{i : \neg A(\mathscr Q(i), \mathscr R(i))} \alpha_i U\ket i
\end{align*}
We know that $\neg A(\mathscr Q(i), \mathscr R(i))\implies \alpha_i = 0$, since $\ket{\psi}$ satisfies $A(q,r)$. Thus,
\begin{align*}
U\ket{\psi} &= \sum_{i : A(\mathscr Q(i), \mathscr R(i))} \alpha_i U\ket i.
\end{align*}
Now, for all $i$ in this superposition, $A'\left(\mathscr Q(i),\mathscr R(i)\right)$ holds due to \eqref{eq:AimplA'} and $U$ acts as the identity, i.e.,
\begin{equation*}
\forall i : A\left(\mathscr Q(i), \mathscr R(i)\right) \implies U\ket i = \ket i,
\end{equation*}
which lets us conclude that
\begin{align*}
U\ket{\psi} &= \sum_{i : A(\mathscr Q(i), \mathscr R(i))} \alpha_i \ket i = \ket{\psi}.
\end{align*}
Removing $U$ from $\mathcal C$ does thus not affect $\ket\psi$ and, in particular, the final outcome of running $\mathcal P$ will not change by performing this optimization. Furthermore, our methodology will not modify the circuit if \eqref{eq:AimplA'} does not hold.
\end{proof}

\section{Formalization and Generalization}
In this section, we formalize the pre- and postconditions that are necessary to handle all examples in this paper, including the practical examples that will be introduced in the next section. Furthermore, we introduce a generalization of our methodology that is strictly more powerful.

\subsection{Formalization of our Basic Methodology}

In order to formalize the basics of our methodology, we first define the pre- and postconditions of the quantum subroutines that are required for our examples in Table~\ref{tbl:prepost}, 
where $X(q)$ denotes application of a Pauli-X~\cite{barenco1995elementary} gate to qubit $q$.

\begin{table}[h]
\begin{tabular}{lll}
	\centering
	Operation & Preconditions & Postconditions\\
	\toprule
	$q$=Alloc($n$) & $\{q=\emptyset; n\in\mathds N\}$ &  $\{q = \ket{0}^{\otimes n}\}$\\
	Dealloc($q$) &$\{q=\ket{0}^{\otimes n}\}$ &  $\{q=\emptyset\}$\\
	 Swap($q_i$,$q_j$) & $\{q_i=A, q_j=B\}$ & $\{\,q_i=B, q_j=A\}$\\
	X($q$) & $\{q=A,A\in\{0,1\}\}$ & $\{\,q = A\oplus 1\}$\\
	\bottomrule
\end{tabular}
\caption{Pre- and postconditions of the quantum subroutines that are required to optimize our examples.}
\label{tbl:prepost}
\end{table}

From the pre- and postconditions of the Swap operation, it is also apparent that a Swap is trivial if $q_i==q_j$; a fact that we already used in Example~\ref{ex:bell}.

In addition to the pre- and postconditions above, we require a formal description of the control modifier, which turns a given quantum subroutine $U$ into its controlled version $U^c$, where $c$ refers to the control qubit. The postcondition corresponding to
\begin{equation*}
	control(U)(c, q)
\end{equation*}
is $\{q = U^c\ket{\psi}\}$, where $U^c$ is $U$ on the subspace where $c=\ket 1$ and $U^c=\mathds 1$ on the subspace where $c=\ket0$, and $\ket{\psi}$ denotes the state of the $q$-register before applying $control(U)$.

The pre- and postconditions of higher-level subroutines may be defined in a similar fashion. However, when combined, the pre-/postconditions above are sufficient for our examples.
E.g., combining the control modifier with the NOT or Pauli X gate allows us to optimize the Bell-pair example where, after an initial Hadamard gate $H$~\cite{barenco1995elementary} on $\ket{00}$, the controlled NOT gate was applied as follows
\begin{equation*}
	H_1\ket0\ket0=\frac 1{\sqrt 2 }(\ket0+\ket1)\ket0\overset{CNOT}{\mapsto}\frac 1{\sqrt 2}(\ket{00}+\ket{11})\;.
\end{equation*}
Since the controlled NOT gate flips the qubit in $\ket0$ if the control qubit is one, we immediately get the postconditions for the two qubits $q_0$ and $q_1$
\begin{equation*}
\{q_1=X^{q_0}\ket{0}\}\implies\{q_1==q_0\}\;,
\end{equation*}
by combining the pre- and postconditions of the control modifier and the Pauli X gate. Together with the triviality condition of the Swap gate acting on two qubits $q_i$ and $q_j$,
\begin{equation*}
	\{q_i == q_j\}\;,
\end{equation*}
we can again remove the Swap gate from the circuit of the Bell-pair example.
Similarly, the pre- and postconditions for CNOT can be used to identify the optimization opportunity in the following example.

\begin{figure*}[t]
\centering
\resizebox{\linewidth}{!}{
\begin{tikzpicture}
\node (orig) at (0,0) {\includegraphics[height=.13\linewidth]{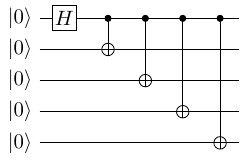}};
\node at (0,-1.8) {\textbf{(a)} Original circuit for entangling all qubits};

\node (orig) at (10,0) {\includegraphics[height=.13\linewidth]{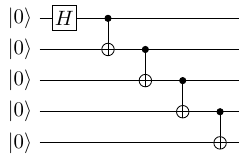}};
\node at (10,-1.8) {\textbf{(c)} Circuit for LNN after optimization.};

\node (orig) at (5,-4.5) {\includegraphics[height=.13\linewidth]{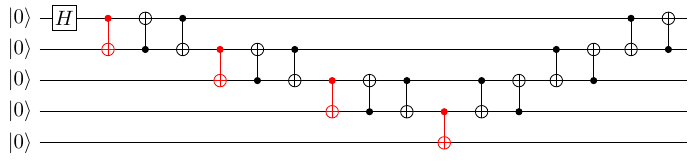}};
\node at (5,-6.2) {\textbf{(b)} Circuit for LNN before A.B. opt.};

\draw[-stealth,very thick,rounded corners=1pt] (0,-2.5)--(0,-4.5) -- (0.9,-4.5);
\draw[-stealth,very thick,rounded corners=1pt] (9.1,-4.5)--(10,-4.5)--(10,-2.5);

\node at (0.3,-3.5) {\rotatebox[]{-90}{1D Mapper}};
\node at (-0.3,-3.5) {\rotatebox[]{-90}{$+$ opt.}};
\node at (9.7,-3.5) {\rotatebox[]{90}{A.B. opt.}};
\end{tikzpicture}}
\caption{Optimizing a chain of CNOTs for a linear nearest-neighbor (LNN) architecture such as the 9-qubit chip by Google~\cite{kelly2015state} by employing the pre- and postconditions of CNOT gates. The benefit of our assertion-based optimization (A.B. opt.) can be seen clearly when comparing the circuits in \textbf{(b)} and \textbf{(c)}: No extra CNOTs due to Swaps~\cite{kutin2007computation} are necessary in \textbf{(c)}, resulting in much lower gate count and circuit depth.}
\label{fig:cnotchainoptimization}
\end{figure*}

\begin{example}{}
In addition to circuit optimizations at the logical level, entanglement description assertions and triviality conditions can be used to optimize the circuit for a specific target architecture. Consider the compilation steps outlined in Fig.~\ref{fig:cnotchainoptimization}. After mapping the circuit in \textbf{(a)} to a linear nearest-neighbor connectivity with additional optimizations to cancel intermediate partial Swap chains results in the circuit \textbf{(b)}. As before, we can employ our assertion-based optimizer to remove trivial CNOT gates using the fact that after each red CNOT gate acting on $q_i$ and $q_{i+1}$, it holds that $q_i==q_{i+1}$. The optimized circuit is shown in Fig.~\ref{fig:cnotchainoptimization}\textbf{(c)}.
\end{example}

\subsection{Generalized Optimization Methodology}

By generalizing the basic methodology above, we can greatly increase its optimization capabilities. So far, our optimizer considers single gates at any given point together with all available postconditions of previously executed subroutines. For each such gate, it then determines whether it can be removed from the circuit without altering its output. The generalized strategy considers multiple gates and checks whether the supplied postconditions allow to deduce that the combined action of these gates is trivial, in which case all of these gates can be removed from the circuit.
In order to properly introduce our generalized methodology, we first require a few definitions.

\begin{definition}{Set of control qubits.}{}
For an instruction $U\ket{q_1,...,q_k}$ acting with a (unitary) gate $U$ on $k$ qubits, a set of qubits $\mathcal S\subset \{q_1,...,q_k\}$ is called a \textit{set of control qubits} if there exists a sequence of Swap gates $s_1,...,s_t$ acting on pairs from $\{q_1,...,q_k\}$ and a unitary $U'$ such that with $S$ denoting the unitary which performs $s_1,...,s_t$, the following three statements hold.
\begin{enumerate}[label=(\arabic*)]
	\item $S U S^\dagger = (\mathds 1- \ket{1\cdots 1}\bra{1\cdots 1})\otimes \mathds 1 + \ket{1\cdots 1}\bra{1\cdots 1}\otimes U'$
	\item the sequence of Swaps $(s_1,...,s_t)$ permutes $(q_1,...,q_k)$ such that the first $|\mathcal S|$ qubits of the resulting tuple are in $\mathcal S$
	\item $\mathcal S$ is the largest such set.
\end{enumerate}
For instructions featuring a non-unitary operation (measurement, allocation, deallocation), the set of control qubits is empty.
\end{definition}

We note that there may be multiple distinct sets of control qubits for a given instruction (as in the following example). For instructions where multiple choices exist, we choose a set of control qubits once and keep it invariant throughout the optimization process.

\begin{example}
As an example of an instruction where multiple choices exist for the set of control qubits, consider the $Z^c$ operation applied to $\ket{q_1 q_0}$, where $Z$ acts with a $(-1)$--phase on $\ket 1$ and leaves $\ket0$ invariant. It is easy to check that
\begin{align*}
	Z^c &= \ket0\bra0\otimes\mathds 1 + \ket1\bra1\otimes Z\\
		&= \mathds 1\otimes\ket0\bra0 + Z \otimes\ket1\bra1\;,
\end{align*}
since for $Z^c$ to be nontrivial, both qubits need to be in $\ket 1$. Either qubit can thus be chosen to be the control qubit and, thus, the set of control qubits is not unique.
\end{example}

\begin{definition}{Target qubit.}{}
A qubit $q$ in an instruction $U\ket{...,q,...}$ is called a \textit{target qubit} if it is not in the set of control qubits of the instruction $U\ket{...,q,...}$.
\end{definition}

\begin{definition}{Target-successive instructions.}{}
Two instructions $I_1,I_2$ with identical target qubits are called \textit{target-successive} if no other instructions are scheduled to be executed between $I_1$ and $I_2$ that involve the target qubits in a way that does not commute with neither $I_1$ nor $I_2$.
\end{definition}

\begin{figure}[t]
\centering
\includegraphics[height=0.25\linewidth]{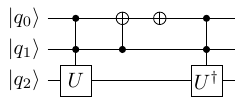}
\caption{Simple example of our multi-gate optimization methodology: The first gate is applied if and only if the last gate is applied (irrespective of the input state $\ket{q_2,q_1,q_0}$). Since the $U$ gate is the inverse of $U^\dagger$, we can cancel the two doubly-controlled gates.}
\label{fig:advtoffoli}
\end{figure}

\begin{figure*}[t]
	\includegraphics[width=\linewidth]{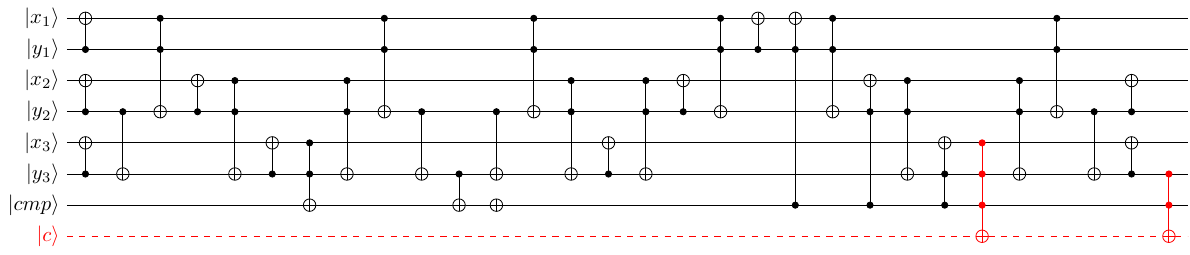}
	\caption{Three-qubit example of a modular adder subroutine which performs the modular reduction. It consists of a comparison, the result of which is stored in the qubit $\ket{cmp}$, and a conditional subtraction. In this setting, our generalized methodology is able to deduce that the two red multi-controlled NOT gates can be canceled, allowing to completely remove the carry qubit $\ket c$. In a modular multiplier, $n$ qubits would be saved since qubit reuse is not generally possible without uncomputation~\cite{bennett73}.}
	\label{fig:cadderopt}
\end{figure*}

Our generalized methodology considers $M\geq1$ target-successive instructions at once, where all $M$ instructions have the same $t$ target qubits and arbitrary controls. Ignoring the control qubits, let $U_1,...,U_M$ denote the $t$-qubit gate matrices of these instructions. An optimization can be performed, for example, if
\begin{equation*}
	U_M\cdots U_1 = \mathds 1_{2^t\times 2^t}
\end{equation*}
and the postconditions on the control qubits are such that either all or none of the gates get executed.
A simple example with $M=2$ and $t=1$ is depicted in Fig.~\ref{fig:advtoffoli}, where the two doubly-controlled gates can be canceled using this reasoning.

We now give a practical example where our multi-gate optimization strategy performs better than the single-gate methodology discussed thus far.

\begin{example}{}\label{ex:modadder}
Consider a circuit that performs addition modulo a quantum number $N$ (stored in another quantum register), i.e.,
\begin{equation*}
	\ket a \ket b\ket N \mapsto \ket{(a+b)\operatorname{ mod }N}\ket b\ket N\;.
\end{equation*}
A possible implementation is to first perform the regular addition, followed by a modular reduction if the result is greater than $\ket N$. Since we only subtract $N$ if $(a+b)\geq N$, the result will always be non-negative and, as a consequence, the final carry qubit will always be zero and it can thus be removed from the subtraction circuit. When using the addition circuit by Takahashi et al.~\cite{takahashi2009quantum}, the optimizer needs to remove the two red multi-controlled NOT gates in Fig.~\ref{fig:cadderopt} which act on the carry qubit in order to exploit this optimization opportunity. Neither of these gates is trivial on its own, but in this setting, either both or none of the two gates are triggered. As a result, this optimization can only be performed using our generalized approach. The achieved reduction in circuit width and depth can be found in Section~\ref{sec:resultsopt}, which discusses the results obtained using our implementation.
\end{example}

\section{Practical examples}\label{sec:examples}

Example~\ref{ex:bell} illustrates that optimization using entanglement description assertions is more powerful than classical constant-folding. Furthermore, Example~\ref{ex:modadder} shows that our generalized methodology is strictly more powerful than regular assertion-based optimization.
In this section, we discuss several practical examples that can be optimized using entanglement description assertions, but not using existing approaches for quantum circuit optimization. We mainly consider subroutines for quantum arithmetic, which is essential for most applications.

Perhaps surprisingly, most of the quantum gates required to run Shor's algorithm for factoring~\cite{shor94} are due to the evaluation of modular exponentiation. In contrast to their classical counterparts, quantum computers must evaluate such classical functions on a superposition of inputs. Because the input is in a superposition, these functions cannot simply be evaluated on a classical computer. This would require reading out the state of the system, which would collapse the superposition and, thus, destroy any quantum speedup. Rather, these functions have to be implemented in terms of quantum gates in order to run them directly on the quantum computer.
Further examples where the evaluation of such classical functions on a quantum computer is necessary are 1) the HHL algorithm for solving linear systems of equations, which requires computing the reciprocal~\cite{harrow2009quantum} and 2) certain algorithms for solving quantum chemistry problems: Babbush et al.~\cite{babbush2016exponentially} have reduced the asymptotic runtime of a chemistry simulation algorithm by computing the entries of the Hamiltonian on-the-fly. This involves evaluating the Coulomb potential and various other mathematical functions which, e.g., describe the chosen orbitals.

In order to enable execution of such classical functions on a quantum computer, one may start by implementing subroutines for basic arithmetic such as addition and multiplication~\cite{haener2018quantum} in terms of quantum gates. These modules can then be combined to enable evaluating polynomials and further higher-level mathematical functions. We use the resulting subroutines as benchmarks and show how our methodology is able to reduce the quantum resource requirements.

\subsection{Floating-Point Arithmetic Subroutine}

\begin{figure}[t]
	\centering
	\includegraphics[width=.6\linewidth]{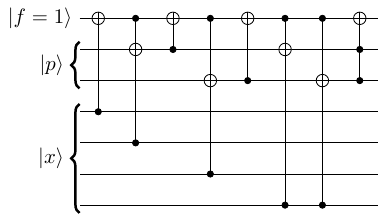}
	\caption{Example of a circuit which finds the first nonzero bit of $\ket x$ and stores its position in $\ket p$ where $\ket x$ is a 4-qubit register and the position register $\ket p$ consists of two qubits~\cite{haener2018quantum}. The flag qubit $\ket f$ is one as long as the first one has not been found.}
	\label{fig:firstone}
\end{figure}

\begin{figure}[t]
	\centering
	\includegraphics[height=.4\linewidth]{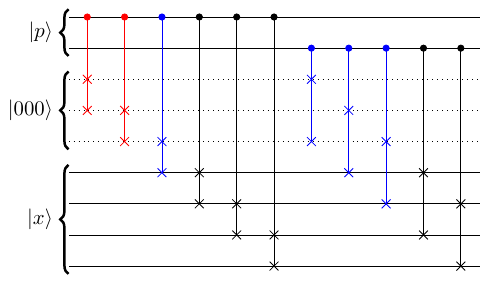}
	\caption{Optimization of the shift circuit that can be performed if $\ket p$ contains the position of the first nonzero bit. Red Fredkin gates can be removed via regular constant-folding. Blue Fredkin gates can be removed using the postconditions of the subroutine depicted in Fig.~\ref{fig:firstone} . As a result, all $2^{n_p}-1$ work qubits can be eliminated (dotted lines).}
	\label{fig:swapoptfigure}
\end{figure}

As a first example in this section, we consider a subroutine that is omnipresent in floating-point arithmetic, namely that of \textit{renormalization}. Renormalization is used during floating-point computations in order to bring intermediate results back into proper floating-point form. This can be achieved using two subroutines: The first subroutine determines the position $p$ of the first nonzero bit of the mantissa. The second subroutine then shifts the mantissa to the left by the output of the first subroutine. A quantum circuit which determines the position of the first nonzero bit is shown in Fig.~\ref{fig:firstone} and a circuit which shifts the mantissa $\ket x$ by $\ket p$ positions is depicted in Fig.~\ref{fig:swapoptfigure}. In order for the shift circuit to work properly for any input, it must allocate $2^{n_p}-1$ extra work qubits in order to catch the overflow from the shifted $\ket x$, where $n_p$ is the number of qubits in the position register $\ket p$. However, in the case where the input to the shift circuit gets initialized by the circuit which determines the position of the first one, such an overflow never occurs. As a result, the $2^{n_p}-1$ work qubits can be eliminated from the combined circuit.

Identifying this optimization opportunity in the complete program is nontrivial and without some description of the action of gates or entire subroutines, such an optimization becomes completely infeasible for large circuits (as it would require simulation thereof for all inputs). We thus introduce a notion of how gates and subroutines interact by providing appropriate entanglement description assertions.

For this concrete example, consider the postcondition of the subroutine which determines the position \texttt{pos} of the first nonzero bit of $\ket x$. It asserts that the first \texttt{pos} qubits of \texttt{x} are zero, i.e.,
\begin{equation*}
	\forall i\in 0..\texttt{pos-1}:\texttt{x[i] == 0}\;,
\end{equation*}
where \texttt{pos} and \texttt{x} are entangled quantum variables.
We can express this equivalently as an {entanglement description assertion} with
\begin{equation*}
	A_{FO}(x,p)=(x < 2^{n-p})\;,
\end{equation*}
where $x$ is interpreted as an integer with $x_0$ as the most-significant bit (MSB) and $p$ corresponds to the position register \texttt{pos} from above with $p_0$ being the least-significant bit (LSB). Using this postcondition, we now optimize the circuit in Fig.~\ref{fig:swapoptfigure}, which achieves the desired shift. Clearly, the red controlled Swap gates -- also known as Fredkin gates -- can be removed since they act on newly allocated qubits which are zero (the postcondition of qubit-allocation says that $q=\ket0^{\otimes n}$). The left-most blue Fredkin gate is a Swap gate controlled on the 0-th bit of $\ket p$ and thus acts trivially if $p_0 = 0$. Furthermore, the Swap itself is trivial if $x_0 = 0$ because all ancilla qubits are still in $\ket 0$. Combining these two triviality conditions of the controlled Swap gate with the postcondition above yields that the blue Fredkin gate may act nontrivially only if
\begin{equation*}
	(p > 0)\land (x < 2^{n-p}) \land (x_0 \neq 0)\;,
\end{equation*}
where $x_0$ denotes the MSB of the $n$-qubit register $x$. Clearly, these conditions cannot hold simultaneously and, as a result, the first blue Fredkin gate in Fig.~\ref{fig:swapoptfigure} can be removed. Combining the postconditions of the Fredkin gates with $A_{FO}(x,p)$ yields a new assertion with
\begin{align*}
	A_{new}(x,p)&=(2^{-p_0}x < 2^{n-p})\\
	&=(x < 2^{n-p+p_0})\;,
\end{align*}
because if the first bit of the position register $p_0$ is one, we have just shifted all of \texttt{x} by one position. Since we successfully removed the first blue Fredkin gate, we can employ regular constant-folding to cancel the second blue Fredkin gate as well (all ancilla qubits are still in $\ket 0$). For the final two blue Fredkin gates, note that they act nontrivially only if
\begin{equation*}
	(p_1 \neq 0)\land ((x_0\neq 0) \lor (x_1\neq 0))\;.
\end{equation*}
From which we can use $p_1\neq 0$ and combine it with the updated postcondition with a case-distinction on $p_0$: If $p_0$ is zero, then $p\geq 2$ and if $p_0$ is one, we have that $p\geq 3$ and that there is a shift of $+1$ in the exponent of the updated postcondition. Thus, in both cases,
\begin{equation*}
	x < 2^{n-2}\;,
\end{equation*}
and hence, the two most-significant bits $x_0,x_1$ of $x$ must be zero.
The action of the remaining two blue Fredkin gates is therefore always trivial and they can also be removed from the circuit.
Finally, since none of the allocated overflow qubits will be used anymore throughout the computation (as their content is always trivial in this application), they will eventually get deallocated without any operations having acted on them. It is then a simple local optimization to cancel allocations with subsequent deallocations, allowing to reduce the width of the resulting circuit by $2^{n_p}-1$ qubits, as desired.

\subsection{Fixed-Point Arithmetic Subroutine}

Similar optimization opportunities arise when using a fixed-point representation. As an example, consider the evaluation of a function using a range reduction, e.g., evaluating the function $f(x)=\sqrt{x}$ for $x\in [0,2)$.

One approach to evaluate $f(x)$ is to approximate the function on the interval $[1,2)$ by a polynomial. We can then perform range reduction for every input $x$ to $y:=x\cdot 2^k\in[1,2)$, for $k\in\mathds N$, evaluate the polynomial for $y$ and then use that
\begin{equation*}
  f(x) = \sqrt{x \cdot 2^k} 2^{-k/2}\;,
\end{equation*}
where both multiplications by powers of two can be implemented using shifts and an additional multiplication by $\sqrt 2$ for the $k/2$ exponent with an odd $k$.
The function $f(x)$ can thus be evaluated on a quantum computer as follows:
\begin{enumerate}[label=\arabic*.]
\item Determine the position of the first non-zero qubit of $x$ (starting from the most-significant bit)
\item Shift all bits of the fixed-point number such that the position is aligned with the binary point (and the number is now in the interval $[1,2)$)
\item Evaluate the polynomial approximating $f(x)$ on $[1,2)$ and store the result in a new quantum register
\item On the result register, undo half of the shift and multiply by $\sqrt 2$ for odd shifts
\end{enumerate}

We know that $x$ was shifted by $k$ positions in step 2. Therefore, the last $k$ qubits of $x$ are zero (where $k$ is a quantum integer). These qubits can be used as work qubits when undoing half of the shift on the result register and our assertion-based optimizer can thus save an entire quantum fixed-point register.

More specifically, after having shifted the qubits of $x$ toward the MSB, we have the same entanglement description assertion as in the previous example, $A_{FO}(x,p)$, between the $x$ register and the position register. Now, the library implementation of the shift circuit should be able to use these ``free'' qubits of $x$ as scratch space when shifting the output of the polynomial evaluation subroutine. This can be achieved through weakening of the preconditions on the work qubits of the shift circuit that catch potential overflow: Instead of requiring $2^{n_p}-1$ qubits in $\ket0$, it is sufficient that the first $k$ qubits be zero, where $k$ is a quantum integer denoting the distance of the shift. While this precondition can always be satisfied by allocating $2^{n_p}-1$ work qubits in $\ket 0$, stating the weakened version allows the compiler to perform this optimization.

\subsection{Integer Addition and Dirty Qubits}

Recently, it was shown that the cost of an integer addition subroutine can be halved using the fact that the target qubit after an uncompute Toffoli gate is back in $\ket0$~\cite{Gidney2018halvingcostof}. Using the pre- and postconditions of allocation and deallocation, our optimization methodology can identify and exploit this optimization opportunity.

A similar optimization can be applied for subroutines that employ so-called \textit{dirty qubits}~\cite{barenco1995elementary,roetteler2017factoring}. While such implementations use fewer (clean) work qubits, they typically cause an increase in circuit depth. Thus, when compiling the program for a specific architecture, the compiler should use as many clean qubits as possible in order to reduce this negative effect on the runtime. Because our assertion-based optimizer is able to identify when a dirty qubit gets mapped to a qubit that is actually clean, it can then optimize the resulting circuit.

For $n$-ary controlled NOT operations~\cite{barenco1995elementary}, this translates to canceling Toffolis that are guaranteed not to be applied because one of the control qubits is in $\ket0$ (since it is a clean qubit). For an addition-by-constant circuit that uses dirty qubits, this translates to removing both the controlled inversions and a controlled incrementer~\cite[Fig. 5]{roetteler2017factoring}. These automatic conversions from dirty to clean qubits allow savings of approximately $2\times$ in the number of gates and, crucially, allow for more modularity when implementing libraries for quantum computing.

As a technical detail, note that gates can be removed both after allocation and before deallocation: By definition, a dirty qubit must be returned to its original state before deallocation and so the clean qubit will have been brought back to $\ket0$.

\section{Implementation using ProjectQ and Z3}
In this section, we discuss our implementation of our optimization methodology. We implement our methodology using the ProjectQ software framework for quantum computing~\cite{steiger2018projectq}. ProjectQ features an extensible compiler framework, allowing to easily integrate custom compiler passes such as our optimization methodology.

For each quantum operation for which we would like to add nontrivial optimization capabilities using our approach, we add the corresponding post- and triviality conditions. Additionally, preconditions may be supplied which would allow to test the program for correctness. To add support for our generalized methodology, we only require the triviality condition of the control modifier, in addition to information which lets us determine whether a sequence of operations $U_1, ..., U_M$ acts as the identity. The latter is already available in ProjectQ.

We extend the definitions of several quantum gates in ProjectQ with the corresponding entanglement descriptions (both post- and triviality conditions). Specifically, we add member functions that use functionality from the Z3 Theorem Prover package~\cite{z3paper} to express these conditions. Our custom compiler pass uses these member functions in combination with the Z3 solver in order to check whether certain operations are guaranteed to be trivial, in which case they can be removed.

While we do not elaborate on the details of the ProjectQ compilation framework, we point out that optimization and compilation is carried out during circuit generation time. As a result, all parameters of the circuit are already known. In particular, the lengths of all quantum registers are known since all classical inputs to the quantum program have been supplied. The circuit can thus be optimized specifically to the problem instance in question. Furthermore, this enables more powerful optimizations when employing our methodology because we do not require parametric proofs. It is of course theoretically possible to prove such statements by induction, however, there is only limited support in automatic theorem provers such as Z3~\cite{z3paper} due to the difficulty of, e.g., constructing appropriate induction rules~\cite{bundy1999automation}. Since all classical parameters have a definite value upon circuit generation, we can unroll quantified statements and thereby generate claims that are easier to prove.

As an example, we show how the definition of the ProjectQ Swap operation was altered in order to enable our optimization engine to carry out the optimizations discussed so far. The definition of \texttt{SwapGate} was extended by merely the following two member functions:
\lstset{tabsize=2,language=python,morekeywords={int,in,return,len,quint,qallocate,deallocate,qureg,qubit},frame=top bottom,breaklines=true,escapeinside={(*}{*)},basicstyle=\small\ttfamily}
\begin{lstlisting}[captionpos=none]
class SwapGate(SelfInverseGate):
    [...]
    def trivial_if(self, x1, x2):
        return (x1 == x2)

    def postconditions(self, x1, x2, y1, y2):
        return And(x1 == y2, x2 == y1)
\end{lstlisting}
Clearly, these are very minor modifications that provide exactly the information required: Postconditions and triviality conditions of the Swap gate. The \texttt{trivial\_if} member function of every gate is invoked by the optimizer with one symbolic boolean variable for each target qubit of the gate (two in this case). The returned expression is negated and then added to the solver together with the expression \texttt{ctrls\_one = And(v[cqb$_1$],\- v[cqb$_2$], ...)}, which is true if and only if all variables \texttt{v[cqb$_i$]} corresponding to control qubits \texttt{cqb$_i$} that are true / equal to one:
\begin{lstlisting}[captionpos=none]
solver.push()
solver.add(And(ctrls_one, Not(cmd.gate.trivial_if((*$\ast$*)target_vars))))
if solver.check() == unsat:
	... # skip current operation
solver.pop()
\end{lstlisting}
where \texttt{target\_vars} are the Z3 variables corresponding to the target qubits of the current gate before it is executed. If the solver finds a solution that satisfies all previous conditions and the negated conditions of \texttt{trivial\_if}, the gate cannot be removed since it may have a nontrivial effect on the state of the quantum computer $\ket{\psi}$ at that point. If there is no such solution, on the other hand, this means that the gate is trivial and it can thus be removed from the circuit.
After this triviality check, the conditions of the Z3 solver are updated according to the postconditions of the operation which hold irrespective of whether the gate was removed: For each target qubit, a new boolean Z3 variable is created and the \texttt{postconditions} member function of the gate relates the old variables (before applying the gate) to the new ones. In particular, operations are handled by adding two Z3 \texttt{Implies(...)} statements: 

\begin{enumerate}
\item The control qubit(s) being all ones implies that the new target variables are now related to the old ones via the \texttt{post\-conditions} function, i.e.,
\begin{lstlisting}[captionpos=none]
Implies(ctrls_one, cmd.gate.postconditions((*$\ast$*)(target_vars+new_target_vars)))
\end{lstlisting}
is added to the solver, where \texttt{new\_target\_vars} are the Z3 variables that correspond to the target qubits after applying the gate.
\item The control qubit(s) not being all ones implies that the new target variables are equal to the old ones, i.e., for all $i$ we add the expression
\begin{lstlisting}[captionpos=none]
Implies(Not(ctrls_one), new_target_vars[i] == target_vars[i]))
\end{lstlisting}
to the solver.
\end{enumerate}
If there are no control qubits, (1) and (2) are of the form
\begin{equation*}
	\{\texttt{true}\implies y=f(x)\}\text{ and } \{\texttt{false}\implies y=x\}\;,
\end{equation*}
respectively and, therefore, are equivalent to stating that $y=f(x)$ holds after the gate has been applied, where $f$ is given by the \texttt{postconditions} member function.
As a technical detail, note that the ProjectQ Swap gate derives from \texttt{Self\-Inverse\-Gate}, stating that the Swap operation is its own inverse. This information is useful for our generalized optimization approach, which is employed whenever the circuit buffer size of the optimizer exceeds a user-defined threshold. When this happens, the stored circuit is traversed in order to identify target-successive operations which may be removed from the circuit. For each such sequence of gates, the Z3 solver is used to determine whether there is an assignment to the control qubits that agrees with all previous postconditions and that causes $0<m<M$ operations to be executed. If there is no such assignment, either all or none of these $M$ operations are executed, meaning that they always act trivially. As a result, the entire sequence of gates can be removed from the circuit.

\section{Results}\label{sec:resultsopt}

In this section, we report the results that were obtained using our prototype implementation of the proposed optimization methodology. We demonstrate the performance of our optimizer using three different quantum subroutines.
The first subroutine performs floating-point mantissa renormalization, see Figs.~\ref{fig:firstone}~and~\ref{fig:swapoptfigure}, the second entangles a linear chain of qubits, see Fig.~\ref{fig:cnotchainoptimization}, and the third performs modular reduction, see Fig.~\ref{fig:cadderopt}, which is a subroutine that is used in constructing a modular adder.

For all subroutines, we compare two ProjectQ compiler setups---one which features a local optimizer capable of merging/canceling subsequent operations that act on the same qubits, and a second configuration which additionally contains our assertion-based optimizer. We choose $\{$CNOT, X, H, S, $T, T^\dagger\}$ as the target gate set for both configurations.
In order to compare these different configurations, we use circuit width and depth as benchmark numbers. The circuit width corresponds to the maximal number of alive qubits at any point throughout the execution of the circuit. The \textit{circuit depth} is equal to the delay of the circuit assuming that all quantum gates in the target gate set take unit time.

\begin{table}[t]
\def\arraystretch{1.5}
\setlength{\tabcolsep}{1em}
\begin{tabular}{lccccc}
\toprule
\textbf n & width & depth & optimized width & optimized depth & area reduction\\
\midrule
8 & 15 & 235 & 7 & 107 & $4.7\times$\\
16 & 47 & 922 & 15 & 388 & $7.4\times$\\
32 & 287 & 5538 & 31 & 1080 & $47.5\times$\\
64 & 2111 & 38903 & 62 & 3229 & $410.2\times$\\
\bottomrule
\end{tabular}
\caption{Optimizer comparison for the floating-point renormalization circuit for 8-, 16-, 32-, and 64-qubit floating-point, where $n_p=3,5,8,11$, respectively, and the mantissa contains $n-n_p-1$ qubits. Our entanglement description based optimizer achieves a reduction in circuit area (width$\times$depth) of up to $410\times$.}
\label{tab:firstcircuitcmp}
\end{table}

\begin{figure}[t]
\centering
\resizebox{0.6\linewidth}{!}{\input{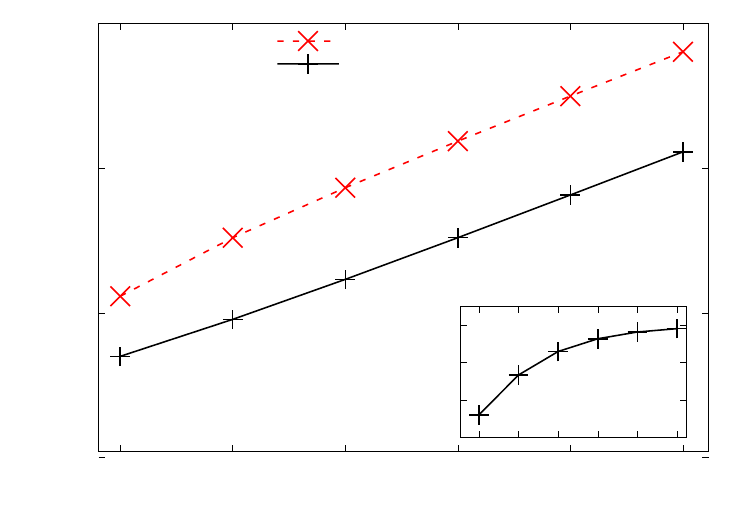}}
\caption{Optimizer comparison for the entangling circuit on $n$ qubits depicted in Fig.~\ref{fig:cnotchainoptimization}. The assertion-based optimizer achieves a $5\times$ improvement in circuit depth for large $n$.}
\label{fig:secondcircuitcmp}
\end{figure}

The comparisons can be found in Table~\ref{tab:firstcircuitcmp} and Fig.~\ref{fig:secondcircuitcmp} for the floating-point renormalization circuit and the entangling circuit, respectively. Both cases clearly demonstrate the benefits of our assertion-based optimizer, which is able to reduce the circuit area (width $\times$ depth) by a factor of up to $410\times$ and $5\times$ for the first and second circuit, respectively.

For the floating-point renormalization circuit, our optimizer is able to eliminate $2^{n_p}-1$ ancilla qubits in addition to several Fredkin gates. Due to the elimination of Fredkin gates, the depth is also significantly reduced. We note that these savings are obtained without any prior manual optimizations such as limiting the maximal shift to the number of bits in the mantissa, as this describes a realistic scenario for software reuse.

For the entangling circuit, all CNOT gates resulting from swap operations can be removed when using the assertion-based optimization strategy (see Example~\ref{ex:bell} for more details). Therefore, the circuit depth would grow by $4(n-2)$ gates for $n\geq 2$ when turning off assertion-based optimization (see Fig.~\ref{fig:cnotchainoptimization}). The ratio between the resulting circuit depths for $n\geq 2$ is thus
\begin{equation*}
	\frac{4(n-2) + n}{n} = \frac{5n-8}{n}\overset{n\rightarrow\infty}{\rightarrow} 5\;,
\end{equation*}
which agrees with the experimental results in Fig.~\ref{fig:secondcircuitcmp} and constitutes an up to $5\times$ improvement over state-of-the-art optimizers.

The modular reduction circuit, which is a subroutine for modular addition, is optimized by identifying a pattern similar (but more complex) to the one shown in Fig.~\ref{fig:advtoffoli}. In this case, the target qubit is the carry qubit of the controlled subtraction and upon removing the two multi-controlled NOT operations, no operations on the carry qubit remain. As a result, our methodology removes this qubit from the circuit. Furthermore, our methodology achieves a slight depth reduction due to the removal of the two generalized Toffoli gates. In Shor's algorithm for factoring an $n$-bit number, $n$ calls to such a modular reduction are required. The total savings would thus be equal to $n$ qubits, where $n\sim 2000$ for practical applications.

\section{Summary and future work}

We have presented an optimization methodology that extends the scope of automatic circuit optimizations. In particular, our methodology carries out high-level optimizations across subroutine boundaries that are typically performed by humans, enabling more efficient code reuse. This is achieved by taking into account pre-, post-, and triviality conditions of all subroutines that get invoked by the quantum program that is being optimized.

Our generalized methodology currently performs optimizations if the overall action of a sequence of gates is trivial. Future work could address more general cases where, e.g., control qubits are in a state that only triggers subsets of these gates that, when combined, correspond to trivial operations. Additionally, symbolic computation on entanglement description assertions may be incorporated. This would allow to optimize iterative procedures such as the Newton-Raphson method which can be used to evaluate high-level arithmetic functions on a quantum computer~\cite{cao2013}: For many such functions, the initial guesses can be chosen to be very simple (e.g., integer powers of two). The first iteration of a Newton-Raphson method may then be applied symbolically to the output of the initial guess routine. Such optimizations have been shown to yield significant resource savings when performed manually~\cite{haner2018optimizing}. Automating such procedures would thus result in the same benefits without the need for labor-intensive manual code optimization.
Moreover, the focus of the present work lies on optimizing permutation-type subroutines. Future work could extend this to target phase-oracles by supporting pre- and postconditions in multiple bases.
\begin{acks}
We thank the anonymous referees for their valuable comments and suggestions.
This work was supported by Microsoft and by the Swiss National Science Foundation through the National Competence Center for Research NCCR QSIT. 
\end{acks}


\bibliography{references}

\end{document}